\documentclass[twocolumn,floatfix,showpacs]{revtex4}
\usepackage{graphicx}
\usepackage{amsmath}
\usepackage{amssymb}
\usepackage{bm}

\begin{document}

\title{Anyonic interferometry without anyons: How a flux qubit can read out a topological qubit}
\author{F. Hassler, A. R. Akhmerov, C.-Y. Hou, and C. W. J. Beenakker}
\affiliation{Instituut-Lorentz, Universiteit Leiden, P.O. Box 9506, 2300 RA Leiden, The Netherlands}
\date{May 2010}
\begin{abstract}
Proposals to measure non-Abelian anyons in a superconductor by quantum interference of vortices suffer from the predominantly classical dynamics of the normal core of an Abrikosov vortex. We show how to avoid this obstruction using coreless Josephson vortices, for which the quantum dynamics has been demonstrated experimentally. The interferometer is a flux qubit in a Josephson junction circuit, which can nondestructively read out a topological qubit stored in a pair of anyons --- even though the Josephson vortices themselves are not anyons. The flux qubit does not couple to intra-vortex excitations, thereby removing the dominant restriction on the operating temperature of anyonic interferometry in superconductors.
\end{abstract}
\maketitle

A topological quantum computer makes use of a nonlocal way of storing quantum information in order to protect it from errors \cite{Kit03,Nay08}. One promising way to realize the nonlocality is to store the information inside the Abrikosov vortices that form when magnetic field lines penetrate a superconductor. Abrikosov vortices can trap quasiparticles within their normal core \cite{Car64}, which in special cases are anyons having non-Abelian statistics \cite{Rea00,Iva01}. For this to happen, the vortex should have a midgap state of zero excitation energy, known as a Majorana bound state. While vortices in a conventional \textit{s}-wave superconductor lack Majorana bound states, they are expected to appear \cite{Vol99,Fu08,Sau10a,Ali10a} in the chiral \textit{p}-wave superconductors that are currently being realized using topological states of matter.

The method of choice to read out a nonlocally encoded qubit is interferometry \cite{Ste06,Bon06}. A mobile anyon is split into a pair of partial waves upon tunneling, which interfere after encircling an even number of stationary anyons. (There is no interference if the number is odd.) The state of the qubit encoded in the stationary anyons can be read out by measuring whether the interference is constructive or destructive. The superconducting implementation of this anyonic interferometry has been analyzed in different setups \cite{Fu09,Akh09,Gro10,Sau10b}, which suffer from one and the same impediment: Abrikosov vortices are massive objects that do not readily tunnel or split into partial waves. 

The mass of an Abrikosov vortex is much larger than the bare electron mass because it traps a large number of quasiparticles. (The enhancement factor is $k_{F}^{3}\xi^{2}d$, with $d$ the thickness of the superconductor along the vortex, $\xi$ the superconducting coherence length, and $k_{F}$ the Fermi wave vector \cite{Vol97}.) There exist other ways to make Majorana bound states in a superconductor (at the end-points of a semiconducting wire or electrostatic line defect \cite{Kit01,Wim10,Lut10,Ore10}), but these also involve intrinsically classical objects. If indeed Majorana bound states and classical motion go hand in hand, it would seem that anyonic interferometry in a superconductor is ruled out --- which would be bad news indeed.

Here we propose an alternative way to perform the interferometric read out, using quantum Josephson vortices instead of classical Abrikosov vortices as the mobile particles. A Josephson vortex is a $2\pi$ twist of the phase of the order parameter, at constant amplitude. Unlike an Abrikosov vortex, a Josephson vortex has no normal core so it does not trap quasiparticles. Its mass is determined by the electrostatic charging energy and is typically less than $1\%$ of the electron mass \cite{Faz01}. Quantum tunneling and interference of Josephson vortices have been demonstrated experimentally \cite{Wal03,Eli93}. This looks promising for anyonic interferometry, but since the Josephson vortex itself is not an anyon (it lacks a Majorana bound state), one might object that we are attempting anyonic interferometry without anyons. Let us see how this can be achieved, essentially by using a non-topological flux qubit \cite{Wal00,Tiw07} to read out the topological qubit.

\begin{figure}[tb]
\centerline{\includegraphics[width=0.6\linewidth]{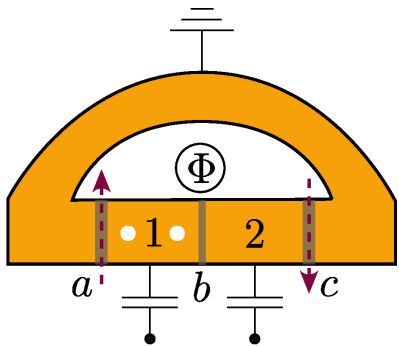}}
\caption{\label{fig:setup}
Circuit of three Josephson junctions $a,b,c$, two superconducting islands $1,2$, and a superconducting ring (enclosing a flux $\Phi$). A persistent current can flow clockwise or counterclockwise. This flux qubit can read out the state of a topological qubit stored in one of the two islands (white discs). Dashed arrows indicate the Josephson vortex tunneling events that couple the two states of the flux qubit, leading to a tunnel splitting that depends on the state of the topological qubit.
}
\end{figure}

We consider a Josephson junction circuit (see Fig.\ \ref{fig:setup}) which can exist in two degenerate states $|L\rangle$, $|R\rangle$, distinguished by the phases $\phi_{i}^{L}$, $\phi_{i}^{R}$ of the order parameter on the islands. The supercurrent flows to the left or to the right in state $|L\rangle$ and $|R\rangle$, so the circuit forms a flux qubit (or persistent current qubit). This is a non-topological qubit. 

The topological qubit is formed by a pair of non-Abelian anyons in a superconducting island, for example the midgap states in the core of a pair of Abrikosov vortices. The two states $|0\rangle$, $|1\rangle$ of the topological qubit are distinguished by the parity of the number $n_{p}$ of particles in the island. For $n_{p}$ odd there is a zero-energy quasiparticle excitation shared by the two midgap states. This qubit is called topological because it is insensitive to local sources of decoherence (since a single vortex cannot tell whether its zero-energy state is filled or empty).

To measure the parity of $n_{p}$, and hence read out the topological qubit, we make use of the suppression of macroscopic quantum tunneling by the Aharonov-Casher (AC) effect \cite{Tiw07,Fri02}. Tunneling from $|L\rangle$ to $|R\rangle$ requires quantum phase slips. If the tunneling can proceed along two path ways, distinguished by a $2\pi$ difference in the value of $\phi_{1}^{R}$, then the difference between the two tunneling paths amounts to the circulation of a Josephson vortex around the island containing the topological qubit (dashed arrows in Fig.\ \ref{fig:setup}).

According to the Aharonov-Casher (AC) effect, a vortex encircling a superconducting island picks up a phase increment $\psi_{\rm AC}=\pi q/e$ determined by the total charge $q$ coupled capacitively to the superconductor \cite{Ste08}. (The charge may be on the superconducting island itself, or on a nearby gate electrode.) If $q$ is an odd multiple of the electron charge $e$, the two tunneling paths interfere destructively, so the tunnel splitting vanishes, while for an even multiple the interference is constructive and the tunnel splitting is maximal. A microwave measurement of the splitting of the flux qubit thus reads out the topological qubit.

Since we only need to distinguish maximal from minimal tunnel splitting, the flux qubit does not need to have a large quality factor (limited by $1/f$ charge noise from the gate electrodes). Moreover, the read out is insensitive to sub-gap excitations in the superconductor --- since these do not change the fermion parity $n_{p}$ and therefore do not couple to the flux qubit. This \textit{parity protection} against sub-gap excitations is the key advantage of flux qubit read-out \cite{Akh10}.

Following Ref.\ \cite{Tiw07} we assume that the ring is sufficiently small that the flux generated by the supercurrent can be neglected, so the enclosed flux $\Phi$ equals the externally applied flux. Junctions $a$ and $c$ are assumed to have the same critical current $I_{\rm crit}$, while junction $b$ has critical current $\alpha I_{\rm crit}$. Because the phase differences across the three junctions $a,b,c$ sum to $\delta\phi_{a}+\delta\phi_{b}+\delta\phi_{c}=2\pi\Phi/\Phi_{0}$ (with $\Phi_{0} = h/2e$ the flux quantum), we may take $\delta\phi_{a}$ and $\delta\phi_{c}$ as independent variables. The charging energy $E_{C}=e^{2}/2C$ of the islands (with capacitance $C$) is assumed to be small compared to the Josephson coupling energy $E_{J} = \Phi_{0} I_{\rm crit} /2 \pi$, to ensure that the phases are good quantum variables. The phase on the ring is pinned by grounding it, while the phases on the islands can change by Josephson vortex tunneling events (quantum phase slips).

\begin{figure}[tb]
\centerline{\includegraphics[width=0.6\linewidth]{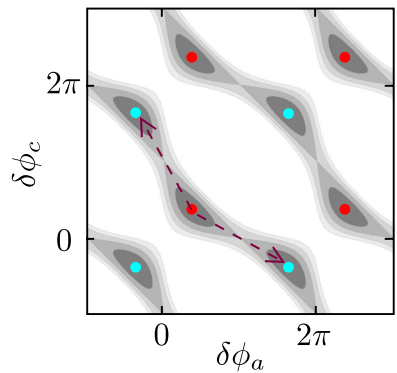}}
\caption{\label{fig:potential}
Contour plot of the potential energy \eqref{eq:josephson} of the flux qubit for $\alpha=1.3$ and $\Phi=\Phi_{0}/2$ (white is high potential, black is low potential). The red and blue dots indicate the minima of clockwise or counterclockwise persistent current. All red dots and all blue dots are equivalent, because the phase differences $\delta\phi_{a},\delta\phi_{c}$ across the Josephson junctions are defined modulo $2\pi$. Tunneling between two inequivalent minima occurs predominantly along the two pathways indicated by the arrows.
}
\end{figure}

The superconducting energy of the ring equals
\begin{align}
U_{J} ={}& -E_{J} [\cos \delta\phi_{a} + \cos \delta\phi_{c}\nonumber \\
&+  \alpha \cos (  2\pi \Phi/\Phi_0 - \delta\phi_{a} - \delta\phi_{c} ) ].
\label{eq:josephson}
\end{align}
The states $|L\rangle$ and $|R\rangle$ correspond in the potential energy landscape of Fig.\ \ref{fig:potential} to the minima indicated by red and blue dots, respectively. Because phases that differ by $2\pi$ are equivalent, all red dots represent equivalent states and so do all blue dots. For $\alpha>1$ the minima are connected by two tunneling paths (arrows), differing by an increment of $+2\pi$ in $\delta\phi_{a}$ and $-2\pi$ in $\delta\phi_{c}$. The difference amounts to the circulation of a Josephson vortex around both islands $1$ and $2$. The two interfering tunneling paths have the same amplitude, because of the left-right symmetry of the circuit. Their phase difference is $\psi_{\rm AC}=\pi q/e$, with $q=\sum_{i=1,2}\bigl(en_{p}^{(i)}+q_{\rm ext}^{(i)}\bigr)$ the total charge on islands and gate capacitors.

The interference produces an oscillatory tunnel splitting of the two levels $\pm\frac{1}{2}\Delta E$ of the flux qubit,
\begin{equation}
\Delta E = E_{\rm tunnel} \bigl|\cos(\psi_{\rm AC}/2) \bigr|.\label{eq:prob}
\end{equation}
Tiwari and Stroud \cite{Tiw07} have calculated $E_{\rm tunnel}\approx 100\,\mu{\rm eV}\simeq 1\,{\rm K}$ for parameter values representative of experimentally realized flux qubits \cite{Wal00} ($E_{J}=800\,\mu {\rm eV}$, $E_{C}=10\,\mu {\rm eV}$). They conclude that the tunnel splitting should be readily observable by microwave absorption at temperatures in the $100\,{\rm mK}$ range.

To read out the topological qubit one would first calibrate the charge $q_{\rm ext}^{(1)}+q_{\rm ext}^{(2)}$ on the two gate capacitors to zero, by maximizing the tunnel splitting in the absence of vortices in the islands. A vortex pair in island $1$ can bind a quasiparticle in the midgap state, allowing for a nonzero $n_{p}^{(1)}$ (while $n_{p}^{(2)}$ remains zero without vortices in island $2$). A measurement of the tunnel splitting then determines the parity of $n_{p}^{(1)}$ (vanishing when $n_{p}^{(1)}$ is odd), and hence reads out the topological qubit.

\begin{figure}[tb]
\centerline{\includegraphics[width=1\linewidth]{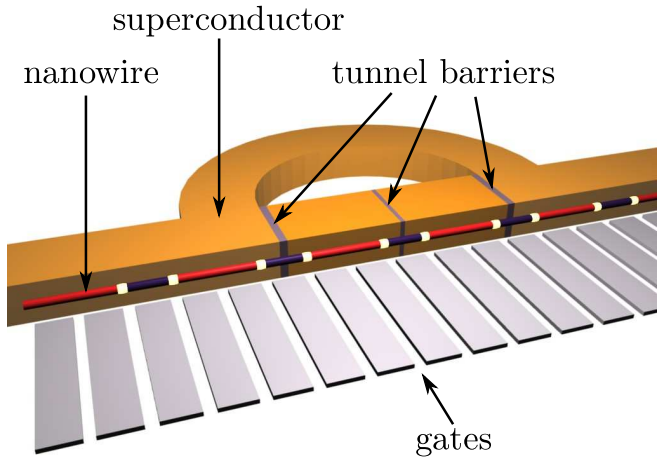}}
\caption{\label{fig_wire}
Register of topological qubits, read out by a flux qubit in a superconducting
ring. The topological qubit is encoded in a pair of Majorana bound states
(white dots) at the interface between a topologically trivial (blue) and
a topologically nontrivial (red) section of an InAs wire. The flux qubit is encoded in the clockwise or counterclockwise persistent current in the ring. Gate electrodes (grey) can be used to move the Majorana bound states along the wire.}
\end{figure}

To implement this read-out scheme the absence of low-energy excitations near the Josephson junction is desirable in order to minimize decoherence of the Josephson vortex as it passes along the junction. The metallic edge states of a topological superconductor are a source of low-energy excitations that one would like to keep away from the junction. So for the flux qubit we would choose a conventional (non-topological) \textit{s}-wave superconductor such as Al or Nb. 

Since a vortex in a non-topological superconductor has no Majorana bound states, we turn to one of the vortex-free alternatives \cite{Kit01,Wim10,Lut10,Ore10}. The ``Majorana wire'' \cite{Lut10,Ore10} seems particularly suitable: A single-mode semiconducting InAs nanowire in a weak (0.1\,{\rm T}) parallel magnetic field is driven into a chiral \textit{p}-wave superconducting state by the interplay of spin-orbit coupling, Zeeman effect, and the proximity to an \textit{s}-wave superconductor. A pair of Majorana bound states is formed at the end points of the wire, provided it is long compared to $\xi$. For that reason Nb ($\xi\lesssim 40\,{\rm nm}$) is to be preferred over Al as superconducting substrate.

A long InAs wire running through a Josephson junction circuit could
conveniently form a register of topological qubits, as illustrated in Fig.\
\ref{fig_wire}. Gate electrodes (grey) deplete sections of the wire (blue)
such that they enter a topologically trivial phase, producing a pair of
Majorana bound states (white dots) at the end points of the topologically
nontrivial sections (red). Each pair encodes one topological qubit,
which can be reversibly moved back and forth along the wire by adjusting,
the gate voltage. (The wire is not interrupted by the tunnel barriers,
of thickness $\ll\xi$.) Once inside the circuit, the tunnel splitting of
the flux qubit measures the state of the topological qubit.

For a universal quantum computation the flux qubit read-out discussed here should be combined with the ability to exchange adjacent Majorana bound states, using two parallel registers \cite{Ali10b}. This is the topologically protected part of the computation. In addition, one needs to perform single-qubit rotations, which as a matter of principle lack topological protection \cite{Nay08}. In the Appendix we show how the flux qubit can be used for parity protected single-qubit rotations (by slowly increasing the flux through the ring from zero  to a value close to $\Phi_{0}/2$ and back to zero).

In comparison with existing read-out schemes \cite{Kit03,Fu08,Fu09,Akh09,Gro10,Sau10b,Sto06}, there are two key differences with the flux qubit read-out proposed here. Firstly, unlike proposals based on the fusion of vortices, our scheme is nondestructive --- meaning that the topological qubit remains available after the measurement (necessary for the realization of a two-qubit {\sc cnot} gate, see the Appendix). 

Secondly, our use of \textit{coreless} vortices to perform the interferometry provides protection against subgap excitations. This parity protection is essential because  the operating temperature would otherwise be restricted to unrealistically small values (below $0.1\,{\rm mK}$ for a typical Abrikosov vortex \cite{Car64}). The characteristic temperature scale for flux qubit read-out is larger by up to three orders of magnitude.

\acknowledgments

We have benefited from discussions with J. E. Mooij. This research was supported by the Dutch Science Foundation NWO/FOM and by an ERC Advanced Investigator Grant. During the final stages of this work, two of us enjoyed the hospitality of the KITPC in Beijing.

\appendix

\section{How a flux qubit enables parity-protected quantum computation with topological qubits}

\subsection{Overview}
\label{sec:overview}

In the main text we discussed the read out of a topological qubit by coupling it to a flux qubit through the Aharonov-Casher effect. This read out is nondestructive (the topological qubit remains available after the read out) and insensitive to subgap excitations (since these do not change the fermion parity). In this Appendix we show, in Sec.\ \ref{sec:cnot}, how flux qubit read-out supplemented by braiding operations \cite{Ali10b} provides the topologically protected part of a quantum computation (in the form of a {\sc cnot} gate acting on a pair of qubits). 

For a universal quantum computer, one needs additionally to be able to perform single qubit rotations of the form
\begin{equation}
|0\rangle+|1\rangle\mapsto e^{-i\theta/2}|0\rangle+e^{i\theta/2}|1\rangle.\label{rotationtheta}
\end{equation}
(Such a rotation over an angle $\theta$ is also called a $\theta/2$ phase gate.) In general (for $\theta$ not equal to a multiple of $\pi/2$), this part of the quantum computation is not topologically protected. A more limited protection against subgap excitations, which do not change the fermion parity, is still possible \cite{Akh10}. We will show in Sec.\ \ref{sec:rotation} how the flux qubit provides a way to perform parity-protected rotations.

In order to make this Appendix self-contained, we first summarize in Sec.\ \ref{sec:background} some background information on topological quantum computation with Majorana fermions \cite{Nay08}. Then we discuss the topologically protected {\sc cnot} gate and the parity-protected single-qubit rotation.

\subsection{Background information}
\label{sec:background}

\subsubsection{Encoding of a qubit in four Majorana fermions}
\label{sec:encoding}

In the main text we considered a qubit formed out of a pair of Majorana bound states. The two states $|0\rangle$ and $|1\rangle$ of this elementary qubit differ by fermion parity, which prevents the creation of a coherent superposition. For a quantum computation we combine two elementary qubits into a single logical qubit, consisting of four Majorana bound states. Without loss of generality we can assume that the joint fermion parity is even. The two states of the logical qubit are then encoded as $|00\rangle$ and $|11\rangle$. These two states have the same fermion parity, so coherent superpositions are allowed.

The four Majorana operators $\gamma_{i}$ ($i=1,2,3,4$) satisfy $\gamma_{i}^{\dagger}=\gamma_{i}$, $\gamma_{i}^{2} =\frac{1}{2}$, and the anticommutation relation $\{\gamma_{i},\gamma_{j}\}=\delta_{ij}$. They can be combined into two complex fermion operators,
\begin{equation}
a_{1} = \frac{\gamma_{1} + i \gamma_{2} }{\sqrt{2}},\;\; a_{2} = \frac{\gamma_{3} + i \gamma_{4} }{\sqrt{2}},\label{agammarel}
\end{equation}
which satisfy $\{a_{i}^{\vphantom{\dagger}},a_{j}^{\dagger}\}=\delta_{ij}$. The fermion parity operator
\begin{equation}
2a_{1}^{\dagger}a_{1}^{\vphantom{\dagger}}-1=2i\gamma_{1}\gamma_{2}\label{parityoperator}
\end{equation}
has eigenvalues $-1$ and $+1$ in states $|0\rangle$ and $|1\rangle$, respectively.

Pauli operators in the computational basis $|00\rangle,|11\rangle$ can be constructed as usual from the $a,a^{\dagger}$ operators, and then expressed in terms of the $\gamma$ operators as follows:
\begin{equation}
\sigma_{x}=-2i\gamma_{2}\gamma_{3},\;\;\sigma_{y}=2i\gamma_{1}\gamma_{3},\;\;
\sigma_{z}=-2i\gamma_{1}\gamma_{2}.\label{Paulidef}
\end{equation}

\subsubsection{Measurement in the computational basis}
\label{sec:measurement}

An arbitrary state $|\psi\rangle$ of the logical qubit has the form
\begin{equation}
|\psi\rangle=\alpha|00\rangle+\beta|11\rangle,\;\;|\alpha|^{2}+|\beta|^{2}=1.\label{psidef}
\end{equation}
A measurement in the computational basis projects $|\psi\rangle$ on the states $|00\rangle$ or $|11\rangle$. This is a fermion parity measurement of one of the two fundamental qubits that encode the logical qubit. 

Referring to the geometry of Fig.\ \ref{fig_wire}, one would perform such a nondestructive projective measurement (called a quantum nondemolition measurement) by moving the Majorana fermions $\gamma_{1},\gamma_{2}$ along the InAs wire into the Josephson junction circuit, while keeping the Majorana fermions $\gamma_{3},\gamma_{4}$ outside of the circuit. Read out of the flux qubit would then measure the fermion parity of the first fundamental qubit, thereby projecting the logical qubit onto the states $|00\rangle$ or $|11\rangle$.

\subsubsection{Braiding of Majorana fermions}
\label{sec:braiding}

The Majorana bound states in the geometry of Fig.\ \ref{fig_wire} are separated by insulating regions on a single InAs wire, so they cannot be exchanged. The exchange of Majorana fermions, called ``braiding'' is needed to demonstrate their non-Abelian statistics. It is also an essential ingredient of a topologically protected quantum computation. In order to be able to exchange the Majorana bound states one can use a second InAs wire, running parallel to the first and connected to it by side branches. Braiding of Majorana fermions in this ``railroad track'' geometry has been studied recently by Alicea \textit{et al.} \cite{Ali10b}. We refer to their paper for the details of this implementation and in the following just assume that adjacent Majorana bound states can be exchanged as needed.

The counterclockwise exchange of Majorana fermions $j<j'$ implements the operator \cite{Rea00,Iva01}
\begin{equation}
\rho_{jj'} =2^{-1/2}(1-2\gamma_{j}\gamma_{j'})= e^{(i\pi/4) (2 i \gamma_{j} \gamma_{j'}) }.\label{rhoijdef}
\end{equation}
In view of Eq.\ \eqref{Paulidef}, braiding can therefore generate the unitary operations $\exp[\pm(i\pi/4)\sigma_{k}]$ ($k=x,y,z$). These $\pi/2$ rotations (or $\pi/4$ phase gates) are the only single-qubit operations that can be generated in a topologically protected way \cite{Nay08}.

\subsection{Topologically protected CNOT gate}
\label{sec:cnot}

The controlled-not ({\sc cnot}) two-qubit gate can be carried out in a topologically protected way by the combination of braiding and fermion parity measurements, along the lines set out by Bravyi and Kitaev \cite{Bra02}.

The computational basis, constructed from the first logical qubit formed by Majorana operators $\gamma_1, \gamma_2, \gamma_3 , \gamma_4$ and the second logical qubit $\gamma_5, \gamma_6, \gamma_7, \gamma_8$, consists of the four states
\begin{equation}
\label{eq:Computational-basis-2-qubits}
|00\rangle|00\rangle,\;\;
|00\rangle|11\rangle,\;\;
|11\rangle|00\rangle,\;\;
|11\rangle|11\rangle.
\end{equation}
The first and second kets represent the first and second logical qubits, respectively, and the two states within each ket represent the two fundamental qubits. In this basis, the {\sc cnot} gate has the matrix form
\begin{equation}
\mbox{\sc cnot}=
\begin{pmatrix}
1 & 0 & 0 & 0
\\
0 &1 & 0 & 0
\\
0 & 0 & 0 & 1
\\
0 & 0 & 1 & 0
\end{pmatrix}.\label{cnotmatrix}
\end{equation}
In words, the second logical qubit (the target) is flipped if the first logical qubit (the control) is in the state $|11\rangle$, otherwise it is left unchanged.

For a topologically protected implementation one needs an extra pair of Majorana fermions $\gamma_{9},\gamma_{10}$ (ancilla's), that can be measured jointly with the Majorana fermions $\gamma_{1},\ldots\gamma_{8}$. The {\sc cnot} gate can be constructed from $\pi/2$ rotations (performed by braiding), together with measurements of the fermion parity operator $(2i\gamma_{i}\gamma_{j})(2i\gamma_{k}\gamma_{l})$ of sets of four Majorana fermions \cite{Bra02}. Because the measurements include Majorana fermions from the computational set $\gamma_{1},\ldots\gamma_{8}$ (not just the ancilla's), it is essential that they are nondestructive.

Referring to Fig.\ \ref{fig_wire}, such a nondestructive joint parity measurement can be performed by moving the four Majorana bound states $i,j,k,l$ into the Josephson junction circuit. (The double wire geometry of Ref.\ \cite{Ali10b} would be used to bring the bound states in the required order.) Read out of the flux qubit then projects the system onto the two eigenstates of $(2i\gamma_{i}\gamma_{j})(2i\gamma_{k}\gamma_{l})$ of definite joint parity.

\subsection{Parity-protected single-qubit rotation}
\label{sec:rotation}

\subsubsection{From topological protection to parity protection}
\label{sec:intro}

There is a relatively small set of unitary operations that one needs in order to be able to perform an arbitrary quantum computation. One needs the {\sc cnot} two-qubit gate, which can be done in a topologically protected way by braiding and read out as discussed in Sec.\ \ref{sec:cnot}. One needs $\pi/2$ single-qubit rotations ($\pi/4$ phase gates), which can also be done with topological protection by braiding (Sec.\ \ref{sec:braiding}). These socalled Clifford gates can be efficiently simulated on a classical computer, and are therefore not sufficient. 

One more gate is needed for a quantum computer, the $\pi/4$ single-qubit rotation ($\pi/8$ phase gate). This operation cannot be performed by braiding and read out --- at least not without changing the topology of the system during the operation \cite{Fre06,Bon10a} and incurring both technological and fundamental obstacles \cite{note1,Ran10}. As an alternative to full topological protection, we propose here a parity-protected $\pi/4$ rotation.

Braiding and read out are topologically protected operations, which means firstly that they are insensitive to local sources of decoherence and secondly that they can be carried out exactly. (As discussed in Sec.\ \ref{sec:braiding}, exchange of two Majorana fermions rotates the qubit by exactly $\pi/2$.) The $\pi/4$ rotation lacks the second benefit of topological protection, so it is an approximate operation, but the first benefit can remain to a large extent if we use a flux qubit to perform the rotation in a parity protected way, insensitive to subgap excitations.

The straightforward approach to single-qubit rotations is partial fusion, which lacks parity protection: One would bring two vortices close together for a short time $t$, and let the tunnel splitting $\delta E$ impose a phase difference $\theta=t\delta E/\hbar$ between the two states $|0\rangle$ and $|1\rangle$. The result is the rotation \eqref{rotationtheta}, but only if the vortices remain in the ground state. The minigap in a vortex core is smaller than the bulk superconducting gap $\Delta_{0}$ by a large factor $k_{F}\xi$, so this is a severe restriction (although there might be ways to increase the minigap \cite{note2,Sau09,Mol10}). An alternative to partial fusion using edge state interferometry has been suggested \cite{Bon10b} in the context of the Moore-Read state of the $\nu=5/2$ quantum Hall effect \cite{Moo91}, where parity protection may be less urgent.

Like the parity-protected read-out discussed in the main text, our parity-protected $\pi/4$ rotation uses the coupling of a flux qubit to the topological qubit. The coupling results from the Aharonov-Casher effect, so it is insensitive to any any other degree of freedom of the topological qubit than its fermion parity. The operation lacks topological protection and is therefore not exact (the rotation angle is not exactly $\pi/4$). It can be combined with the distillation protocol of Bravyi and Kitaev \cite{Bra05,Bra06}, which allows for error correction with a relatively large tolerance (error rates as large as 10\% are permitted).

\subsubsection{Method}
\label{sec:method}

As explained in Sec.\ \ref{sec:encoding}, we start from a logical qubit encoded as $|00\rangle$, $|11\rangle$ in the four Majorana fermions $\gamma_{1},\gamma_{2},\gamma_{3},\gamma_{4}$. We bring the Majorana bound states 1 and 2 into the Josephson junction circuit, keeping 3 and 4 outside. The effective Hamiltonian of the Josephson junction circuit is
\begin{equation}\label{eq:ham}
  H = -\tfrac{1}{2}\varepsilon\, \tau_{z} +\tfrac{1}{2} \Delta E\, \tau_{x},
\end{equation}
with energy levels
\begin{equation}
E_{\pm}=\pm\tfrac{1}{2}\sqrt{\varepsilon^{2}+\Delta E^{2}}.\label{Epm}
\end{equation}
The Pauli matrices $\tau_{i}$ act on the two states $|L\rangle$, $|R\rangle$ of the flux qubit (states of clockwise and counterclockwise circulating
persistent current). In the absence of tunneling between these two states, their energy difference $\varepsilon =\varepsilon_{0} (\Phi/\Phi_0-1/2)$ (with $\varepsilon_{0} = 4 \pi E_J \sqrt{1-1/4\alpha^2}$) vanishes when the flux $\Phi$ through the ring equals half a flux quantum $\Phi_{0}=h/2e$. Tunneling leads to a splitting $\Delta E$ given by Eq.\ \eqref{eq:prob}.

Parity protection means that the Majorana bound states 1 and 2 appear in $H$ only through their fermion parity $n_{p}$, which determines $\Delta E=\Delta E(n_{p})$ through the Aharonov-Casher phase. Subgap excitations preserve fermion parity, so they do not enter into $H$ and cannot cause errors.

To perform the single-qubit rotation, we start at time $t=0$ from a flux $\Phi$ far from $\Phi_{0}/2$, when $|\varepsilon|\gg\Delta E$. Then the state $|L\rangle$ is the ground state of the flux qubit and the coupling to the topological qubit is switched off. The flux $\Phi(t)$ is changed slowly to values close to $\Phi_{0}/2$ at $t=t_{f}/2$ and then brought back to its initial value at time $t=t_{f}$. The variation of $\Phi$ should be sufficiently slow (adiabatic) that the flux qubit remains in the ground state, so its final state is $|L\rangle$ times a dynamical phase $e^{i\varphi(n_{p})}$ dependent on the fermion parity of the first of the two topological qubits that encode the logical qubit.
 
 The initial state $|\Psi_{i}\rangle=(\alpha|00\rangle+\beta|11\rangle)|L\rangle$ of flux qubit and logical qubit is therefore transformed into
\begin{equation}
|\Psi_{i}\rangle\mapsto |\Psi_{f}\rangle= \bigl(e^{i\varphi(0)}\alpha|00\rangle+e^{i\varphi(1)}\beta|11\rangle\bigr)|L\rangle.\label{Psif}
\end{equation}
By adjusting the variation of $\Phi(t)$ we can ensure that $\varphi(1)-\varphi(0)=\pi/8$, thereby realizing the desired $\pi/4$ rotation.

\subsubsection{Example}
\label{sec:example}

As an example, we vary the flux linearly in time according to 
\begin{align}
&\frac{\Phi(t)}{\Phi_{0}}-\frac{1}{2} =- \frac{E_0 + \lambda | t
  -t_f/2 |}{\varepsilon_0},\label{eq:phit}\\
&\Rightarrow E_{\pm}=\pm\tfrac{1}{2}\sqrt{(E_0 + \lambda | t -t_f/2|)^{2}+\Delta E^{2}}.\label{Epm2}
\end{align}
We assume $q_{\rm ext}=0$, so $\Delta E(1)=0$ and $\Delta E(0)=E_{\rm tunnel}$. We take $E_{0}\gg E_{\rm tunnel}$, for weak coupling between flux qubit and topological qubit. The condition for the adiabatic approximation \cite{Lan77} then takes the form
\begin{equation}\label{eq:eta}
  \left|\frac{\hbar}{2E_{-}^{2}}\frac{ d E_{-}}{dt}
  \right|_{t=t_{f}/2}\approx\frac{\hbar\lambda}{E_{0}^{2}}\ll 1.
\end{equation}

From time $t=0$ to $t=t_{f}$, the flux qubit accumulates the dynamical phase factor
\begin{equation}
  \varphi(n_{p})=\hbar^{-1}\int_{0}^{t_{f}}dt\, E_{-}(t,n_{p}).\label{phiadiabatic}
\end{equation}
To leading order in the small parameter $E_{\rm tunnel}/E_{0}$ we find
\begin{equation}\label{eq:phase}
  \phi(1) - \phi(0) = \frac{ E_{\rm tunnel}^2} {2 \hbar \lambda} 
        \ln (1 + \lambda t_f / 2E_0).
\end{equation}
By choosing
\begin{equation}\label{eq:tf}
 t_f = \frac{2E_{0}}{\lambda} \bigl[ \exp\bigl(\tfrac{1}{4}\pi \hbar
 \lambda/E_{\rm tunnel}^{2}\bigr) -1 \bigr]  
\end{equation}
we implement a $\pi/4$ rotation. 

In order to maximally decouple the flux qubit from the topological qubit at the start and at the end of the operation, we take $\Phi(t)=0$ at $t=0$ and $t=t_{f}$. In view of Eq.\ \eqref{eq:phit}, this requires $\lambda t_{f}=\varepsilon_{0}-2E_{0}$. Substitution into Eq.\ \eqref{eq:tf} gives the desired optimal value of $\lambda$,
\begin{equation}
\lambda_{\rm opt}=(4/\pi\hbar)E_{\rm tunnel}^{2}\ln(\varepsilon_{0}/2E_{0}),\label{lambdaopt}
\end{equation}
still consistent with the adiabaticity requirement \eqref{eq:eta}. For $E_{\rm tunnel}\ll E_{0}\ll\varepsilon_{0}$ the entire operation then has a duration of order $\hbar\varepsilon_{0}/E_{\rm tunnel}^{2}$, up to a logarithmic factor. The quality factor of the flux qubit should thus be larger than $(\varepsilon_{0}/E_{\rm tunnel})^{2}\simeq E_{J}/E_{C}$ (typically $\simeq 10^{2}$).

\end{document}